\documentclass[amsmath,amssymb,prd,nofootinbib,10pt,superscriptaddress,notitlepage]{revtex4-1}

\usepackage[english]{babel}
\usepackage{amsmath}
\usepackage{graphicx, xcolor}
\usepackage[colorlinks=true, allcolors=blue]{hyperref}
\usepackage{ulem}

\renewcommand{\bm}[1]{{\hbox{\boldmath{$#1$}}}}
\newcommand{\bp}{{\hbox{\boldmath $p$}}}
\newcommand{\bpi}{{\hbox{\boldmath $\pi$}}}
\newcommand{\bphi}{{\hbox{\boldmath $\phi$}}}
\newcommand{\bbeta}{{\hbox{\boldmath $\beta$}}}

\newcommand{\btheta}{{\hbox{\boldmath $\beta$}_{\rm nad}}}

\newcommand{\bx}{{\hbox{\boldmath $x$}}}
\newcommand{\bX}{{\hbox{\boldmath $X$}}}
\newcommand{\bk}{{\hbox{\boldmath $k$}}}
\newcommand{\sbx}{{\hbox{\boldmath\scriptsize $x$}}}
\newcommand{\sbk}{{\hbox{\boldmath\scriptsize $k$}}}
\newcommand{\sbp}{{\hbox{\boldmath\scriptsize $p$}}}

\newcommand{\sbtheta}{{\hbox{\boldmath\scriptsize $\beta$}_{\rm nad}}}
\newcommand{\sbbeta}{{\hbox{\boldmath\scriptsize $\beta$}}}

\newcommand{\bre}[1]{\left\langle #1 \right\rangle}
\newcommand{\ket}[1]{\Bigl\vert #1 \Bigr\rangle}
\newcommand{\bra}[1]{\Bigl\langle #1 \Bigr\vert}
\newcommand{\bket}[1]{\Bigl\vert #1 \Bigr\rangle\!\!\!\Bigr\rangle}
\newcommand{\bbra}[1]{\Bigl\langle\!\!\!\Bigl\langle #1 \Bigr\vert}

\newcommand{\sbm}[1]{{\hbox{\boldmath{\scriptsize$#1$}}}}

\newcommand{\wamc}{{\zeta}_{\alpha}}
\newcommand{\bphiSc}{{\bbeta_\alpha}}

\newcommand{\wam}{{{\hat\zeta}_{\alpha}(t)}}
\newcommand{\bphiS}{{\bphi}_{\alpha}(t)}

\newcommand{\wamvec}{\hat{\vec{\zeta}}}
\newcommand{\pizetavec}{\hat{\vec{\pi}}_\zeta}
\newcommand{\bphiScvec}{{\vec\bbeta}}
\newcommand{\wamcs}{{\zeta}_{\sbm{k}_\rs}}
\newcommand{\wamk}{{\hat\zeta}^{\rs}_{\sbm{k}}}
\newcommand{\wammk}{{\zeta}^{\rs}_{\sbm{k}'}}
\newcommand{\psiH}{\ket{\Psi;  \bphiScvec}}
\newcommand{\psiHlocal}{\ket{\Psi_\alpha;  \bphiSc}}
\newcommand{\psiHlocalbra}{\bra{\Psi_\alpha;  \bphiSc}}

\newcommand{\psiHlocaltilde}{\ket{\Psi_\alpha; \tilde{\bbeta}_\alpha}}
\newcommand{\psiHk}{|\Psi;  \wamcs \rangle}
\newcommand{\rs}{{\rm S}}
\newcommand{\rh}{{\rm H}}
\newcommand{\giop}{{{}^{\raisebox{0.4ex}{{\scriptsize$g$}}}\!\hat{\cal O}}}
\newcommand{\lgopbre}{\bre{{}^{\raisebox{0.4ex}{{\scriptsize$g$}}}\!\hat{\cal O}_\alpha}_{\sbm{\beta}_\alpha} 
}

\newcommand{\lgop}{{{}^{\raisebox{0.4ex}{{\scriptsize$g$}}}\!\hat{\cal O}_\alpha}}

\newcommand{\Leex}{{\bx\cdot\partial_\sbx}}
\usepackage{comment}

\begin{document}

\title{Locality in effective field theory for inflationary soft modes
}

\author{Takahiro Tanaka,}
\email{t.tanaka@tap.scphys.kyoto-u.ac.jp}
\affiliation{Department of Physics, Kyoto University, Kyoto 606-8502, Japan}
\affiliation{Center for Gravitational Physics and Quantum Information, Yukawa Institute for Theoretical Physics, Kyoto University, Kyoto 606-8502, Japan}
\author{Yuko Urakawa}
\email{yukour@post.kek.jp}
\affiliation{Institute of Particle and Nuclear Studies, High Energy Accelerator Research Organization (KEK), Oho 1-1, Tsukuba 305-0801, Japan}
\affiliation{The Graduate University for Advanced Studies (SOKENDAI), Tsukuba 305-0801, Japan}

\begin{abstract}
The gradient expansion and the separate universe approach provide an effective description of inflationary soft modes after coarse-graining shorter-wavelength degrees of freedom.  We formulate a locality condition on the quantum state, requiring that the hard-mode state in each local universe depend on the soft modes only through the local soft-mode values in the same patch.  When this condition is satisfied, the coarse-grained soft-mode dynamics remains local, and loop corrections from hard modes to superhorizon correlators of the adiabatic curvature perturbation are perturbatively suppressed.  This provides a model-independent diagnosis of when enhanced corrections due to hard modes can invalidate the gradient expansion.  We further show that the same locality condition implies a generalized soft theorem, from which the standard consistency relations follow under additional assumptions.  This formulation clarifies the origin of possible deviations from the standard consistency relations in multi-field systems or in a non-attractor background.  We also show that the locality condition guarantees the absence of infrared divergences for the correlators of operators invariant under a large gauge transformation.  Thus, locality of the hard-mode state provides a unified criterion for the effective description of inflationary soft modes, generalized soft theorems, the suppression of hard-mode loop corrections, and the infrared regularity of observable correlators.
\end{abstract}

\maketitle

\section{Introduction}

The gradient expansion~\cite{
Lifshitz:1963ps, Starobinsky:1982ee, Deruelle:1994iz, Salopek:1990jq, Sasaki:1995aw} has long provided a powerful framework for describing the dynamics of the early universe on sufficiently large scales.  When the wavelength of interest is much longer than the scale over which causal microphysics can operate, {\it i.e.}, the Hubble scale in simple models, the ratio between these two scales gives a natural expansion parameter.  This leads to an expansion scheme that is distinct from the standard perturbative expansion in the amplitude of fluctuations.  At leading order in the gradient expansion, spatial regions separated by more than the causal scale evolve independently. This observation underlies the separate universe approach~\cite{Salopek:1990jq, Wands:2000dp, Lyth:2004gb}, in which an inhomogeneous universe is described as a collection of locally homogeneous and isotropic universes with different initial conditions.  The evolution of long-wavelength perturbations can then be obtained by solving ordinary background equations in each local patch, rather than the full set of partial differential equations.  This idea is the basis of the $\delta N$ formalism~\cite{Starobinsky:1982ee, Starobinsky:1986fxa, Sasaki:1995aw, Lyth:2004gb, Sasaki:1998ug}, which has been widely used to compute the evolution of the primordial adiabatic curvature perturbation $\zeta$ and its non-Gaussian spectra.  An application of the $\delta N$ formalism beyond the gradient expansion was recently discussed in Ref.~\cite{Saha:2026cay} by using the classical lattice simulation. 

The gradient expansion is a prescription based on an effective field theory. It starts with coarse-graining shorter wavelength modes than the scale of the causal propagation, {\it i.e.}, the subhorizon modes. For a local theory, different local regions in the universe evolve independently, while the mutual interactions among them are suppressed as the higher orders of the gradient expansion, validating the separate universe approach. In Ref.~\cite{Tanaka:2021dww}, the authors showed that the separate universe approach is applicable when the coarse-grained action is invariant under spatial diffeomorphism and is local, clarifying the concrete condition of locality. Based on this, the $\delta N$ formalism, whose original form was applicable only to compute scalar perturbations such as the adiabatic curvature perturbation in a scalar field system, was generalized to compute arbitrary cosmological fluctuations sourced by fluctuations of fields with arbitrary integer spins~\cite{Tanaka:2023gul} at leading order of the gradient expansion. This enables us to compute, {\it e.g.}, the primordial gravitational waves sourced by gauge fields~\cite{Tanaka:2024mzw}.

As long as the gradient expansion holds, we can compute the dynamics of long-wavelength modes (soft modes) without directly solving the coarse-grained degrees of freedom (hard modes) as usual in the case of a valid effective field theory. However, the coarse-grained theory for the soft modes should be local to maintain the validity of the gradient expansion, which requires some conditions on hard modes. In this paper, we provide the required locality condition that the quantum state of the hard modes must satisfy ($\to$ Eq.~(\ref{eq:locality})). When this locality condition holds, one can show that coarse-graining hard modes do not yield any non-local contributions which break the locality of the coarse-grained action required for the validity of the separate universe approach~\cite{Tanaka:2021dww}. In the last several years, the validity of the gradient expansion was questioned in view of the possibility that enhanced loop effects of hard modes may sizably alter the soft modes, such as those on CMB scales~\cite{Kristiano:2022maq, Kristiano:2023scm} (see also Ref.~\cite{Ota:2022xni} for a relevant discussion about gravitational waves). As pointed out, {\it e.g.}, in Refs.~\cite{Firouzjahi:2023aum, Kawaguchi:2024nnu, Fumagalli:2024oxi}, a lengthy computation of loops can show a cancellation among the contributions from different diagrams, but such computations can obscure the underlying cancellation mechanism. In this paper, to keep the generality of the discussion, we do not assume any concrete model of inflation. Instead, our discussion provides a general diagnosis by clarifying the condition that prohibits such a large enhancement that invalidates the gradient expansion. 

In this paper, the locality condition to be imposed on hard modes (\ref{eq:locality}) will be formulated as a condition imposed on quantum states of the hard modes. It may sometimes be more tractable to rephrase it as a condition on correlators. We will show that the locality condition can be generally rephrased as the so-called soft theorem, which relates a correlator containing soft modes to the one obtained by removing the soft modes from it. We also derive a generalized soft theorem in the presence of multiple light fields or when the background deviates from the attractor trajectory.  

Another related issue, which is often discussed separately, is the infrared (IR) divergence of the loop corrections to primordial perturbations.  A naive computation of infrared loop corrections is known to give rise to IR divergences~\cite{Tsamis:1993ub, Tsamis:1996qq} (see Ref.~\cite{Seery:2010kh} for a review about related issues). In particular, the constant part of $\zeta$ is identified as a mode corresponding to a large gauge transformation, and gauge-dependent quantities can retain an apparent sensitivity to the unobservable value of $\zeta$ averaged over a large spatial volume. The divergent contribution was shown to cancel for operators invariant under large gauge transformations (LGTs), when one chooses, {\it e.g.}, the Euclidean vacuum, which is a non-linear extension of the adiabatic vacuum prescription. This was first addressed by performing a gauge fixing in Refs.~\cite{Urakawa:2009my, Urakawa:2009gb} and subsequently by using an LGT-invariant operator~\cite{Urakawa:2010it, Urakawa:2010kr, Tanaka:2013xe, Tanaka:2017nff} or related symmetry-based cancellations~\cite{Gerstenlauer:2011ti, Senatore:2012nq} (see also Refs.~\cite{Giddings:2010nc, Giddings:2011zd} for a relevant discussion and \cite{ Tanaka:2013caa} for a review). In this paper, along the lines of Ref.~\cite{Tanaka:2017nff}, we show that IR divergences in correlators of LGT-invariant operators are generically absent when the quantum state of the universe satisfies the locality condition~\eqref{eq:locality}.

This paper is organized as follows.  In Sec.~\ref{Sec:IRoverview}, after summarizing our prescription, we introduce the locality condition for hard modes, which ensures various infrared properties discussed above. We also show how this condition implies the existence of Weinberg's adiabatic mode at the operator level.  In Sec.~\ref{Sec:ST}, we derive a generalized soft theorem from the locality condition.  We clarify how the usual consistency relation is recovered by additionally imposing the separability condition for the wave function of soft curvature perturbations.  In Sec.~\ref{Sec:Loops}, we show that loop corrections from hard modes to superhorizon correlators of the adiabatic curvature perturbation are perturbatively suppressed.  In Sec.~\ref{Sec:IRreg}, we revisit the IR regularity of inflationary correlators and show that the absence of IR divergences follows from the same locality condition, and that we can construct quantum states that satisfy the locality condition order by order in perturbation. 

\section{Infrared physics in cosmology}  \label{Sec:IRoverview}
To maintain the generality of the discussion as much as possible, we do not specify a concrete model of inflation. 
We assume that the model under consideration is invariant under the spatial diffeomorphism, especially the dilatation. 
For notational brevity, in what follows, we only consider a model which includes $(N+1)$ scalar fields, $\phi^I$ with $I=0, 1, \cdots,\, N$, as matter fields. However, a generalization to include fields with integer spins proceeds straightforwardly by following the discussion in Ref.~\cite{Tanaka:2017nff}. We denote the conjugate momenta of $\phi^I$ by $\pi_I$, which satisfy the commutation relations
\begin{align}
 [\hat\phi^I(t,\bx),\hat\pi_J(t,\bx')]=ie^{-3\rho}\delta^3(\bx-\bx') \delta^I_J,
\end{align}
where $e^\rho$ is the (background) scale factor and $\rho$ is used as the time coordinate. We use ``$~\hat{}~$'' to represent a quantum operator. 

\subsection{Preliminaries}
In a spatial diffeomorphism invariant theory, the curvature perturbation $\hat\zeta(t,\, \bm{x})$ appears in the combination $e^{\hat\zeta(t,\sbm{x})}dx$. Under the spatial dilatation $x^i \to x^i_\lambda = e^{\lambda} x^i$ with $\lambda$ being constant, $\hat\zeta$ transforms as $\hat\zeta(t,\, \bm{x}) = \hat\zeta_{\lambda}(t,\, \bm{x}_\lambda) + \lambda$. Similarly, the other fields transform as 
\begin{align}
    \hat\phi^I(t,\, \bm{x}) = \hat\phi^I_{\lambda}(t,\, \bm{x}_\lambda) \qquad \qquad (I=0, 1,2,\cdots,N)\,. 
\end{align}
With the identification $\hat\phi^0\equiv \hat\zeta$, the transformation law can be written as 
\begin{align}
  \hat\phi^I_{\lambda}(t,\, \bm{x}) = \hat\phi^I(t,\, \bm{x})+ \Delta_\lambda \hat\phi^I (t,\, \bm{x}) + \hat{\cal O}(\lambda^2)\,,
\end{align}
with 
\begin{align}
  \Delta_\lambda \hat\phi^I (t,\, \bm{x}) =  - \lambda \left(\delta^I_0 + 
 \Leex 
  \hat\phi^I(t,\, \bm{x})\right)\,.  
\end{align}

Let us introduce the generator for the dilatation transformation as 
\begin{align}
  \hat Q \equiv e^{3\rho (t)} \int d^3 \bm{x} \, 
  \Delta_\lambda \hat\bphi(t,\,\bm{x} ) \cdot \hat{\bm{\pi}}(t,\, \bm{x} )  \equiv \int d^3\! \bx\, \hat q(t,\,\bx)\,.  
  \label{Eq:charge_dilatation}
\end{align}
Here, $\hat\bphi\equiv\{\hat\phi^I\}$ represents all fields including $\hat\zeta\equiv\hat\phi^0$ collectively, and its conjugate momenta and its linear-order transformation under the dilatation transformation are denoted by $\hat{\bm{\pi}} (t,\, \bm{x} )$ and $\Delta_\lambda \hat\bphi(t,\,\bm{x})$, respectively. We also write $\hat{\pi}^0$ as $\hat{\pi}_\zeta$ interchangeably. The generator $\hat Q$ should commute with the total Hamiltonian $\hat H$, because it is merely a generator of a global spatial coordinate transformation. 
Hence, although the above expression for $\hat Q$ looks time-dependent, $\hat Q$ is actually independent of $t$. 
The commutators between $\hat Q$ and the fields provide, {\it e.g.},  
\begin{align}
 & [i \hat Q,\, \hat \bphi(t,\, \bm{x})] = \Delta_\lambda \hat \bphi (t,\, \bm{x}) \,, \quad
[i \hat Q,\, \hat{\bm{\pi}}(t,\, \bm{x})] = -\lambda\,\frac{\partial}{\partial \bx}\cdot \bx\,
  \hat{\bm{\pi}} (t,\, \bm{x}) \,.
  \label{Eq:commutator_dilatation}
\end{align}
The latter relation can be rewritten as 
\begin{align}
  [i \hat Q,\, e^{-3\hat \zeta}\hat{\bm{\pi}}(t,\, \bm{x})] = -\lambda\,\Leex
  (e^{-3\hat \zeta}\hat{\bm{\pi}} (t,\, \bm{x})) \,,
\end{align}
which means that $e^{-3\hat\zeta}\hat \pi_I$ is a quantity which transforms as a scalar. 
The quantities including spatial derivative in the combination of $e^{-\hat\zeta}\partial_{\sbx}$ also transform in the same way as 
\begin{align}
& [i \hat Q,\, e^{-\hat \zeta(t,\, \sbm{x})}\partial_i\hat \bphi(t,\, \bm{x})]
     = - \lambda \Leex
     \left(e^{-\hat\zeta(t,\, \sbm{x})}\partial_i \hat\bphi(t,\, \bm{x})\right). 
\end{align}

Next, we discuss the action of the dilatation in a local patch. 
Let us denote local patches by ${\cal U}_\alpha$ with $\alpha=1,\ 2,\, 3,\, \cdots$, which cover the entire time slice as a whole. 
The size of each local patch is set to be larger than the horizon scale already at the initial time. 
In this paper, we fix the spatial domain of each local patch in comoving coordinates, and hence its physical volume increases in time. 

We define the spatial coarse-grained fields 
averaged over each local patch ${\cal U}_\alpha$ as
\begin{align}
  \hat\bphi_\alpha(t) \equiv \hat V_\alpha^{-1} \int_{{\cal U}_\alpha} d^3\! \bm{x}\, e^{3\hat\zeta}\hat\bphi (t,\, \bm{x}) \,,
  \qquad
  \hat\bpi_\alpha(t) \equiv L^{-3} \int_{{\cal U}_\alpha} d^3\! \bm{x}\, \hat\bpi (t,\, \bm{x}) \,,
\end{align}
with 
\begin{align}
 \hat V_\alpha \equiv\int_{{\cal U}_\alpha} d^3\! \bm{x}\, e^{3\hat\zeta}\,. 
\end{align}
We express the coarse-grained spatial coordinates as ${\bm{x}}_\alpha$, which takes discrete values, labeling the local patch ${\cal U}_\alpha$. 
Here, we choose the comoving volume of each local patch to be $L^3$, where $L$ denotes a comoving scale of interest, {\it e.g.}, the CMB scale. As is clear from the definition, these coarse-grained fields are composed of soft modes $k \equiv |\bm{k}|$ which satisfies $k< \pi/L$. We denote the set $\{\hat\bphi_\alpha\}$ and $\{\hat\bpi_\alpha\}$ by $\hat{\vec\bphi}$ and $\hat{\vec\bpi}$, respectively. Suppressing the overlapping regions among different local patches, one can choose the coarse-grained fields so that they satisfy
\begin{align}
    \left[ \hat\phi^I_\alpha(t),\, \hat\pi_{J \beta}(t) \right] = i e^{- 3 \rho}L^{-3}  \delta_{\alpha,\, \beta} {\delta^I}_J\,. 
\end{align}

We define the hard modes in $\,{\cal U}_\alpha$ as
\begin{align}
  \hat\bphi{}_{\alpha}^{\rm H}(t,\, \bm{x}) 
  \equiv\hat\bphi(t,\, \bm{x}) - \hat\bphi_\alpha\,,
\qquad
  (e^{-3\hat \zeta} \hat\bpi{}_{\alpha})^{\rm H}(t,\, \bm{x}) 
  \equiv e^{-3\hat \zeta} \hat\bpi(t,\, \bm{x}) - L^3 \hat V^{-1}_\alpha \hat\bpi_\alpha\,,
\end{align}
for $\bx \in {\cal U}_{\alpha}$. 
Using Eqs.~\eqref{Eq:commutator_dilatation}, we obtain
\begin{align}
  & [i \hat Q,\, \hat\phi^{I}_\alpha(t)] = 
    - \lambda \left[\delta^I_0
     +V^{-1}_\alpha \int_{{\cal U}_\alpha} d^3\! \bm{x}\, \partial_\sbx\cdot
     \left(\bx\, e^{3\hat\zeta}\left(\hat\phi^I-\hat\phi^I_\alpha\right) \right) \right]
    \simeq  - \lambda \delta^I_0\, 
    \,,\cr
& [i \hat Q,\, L^3 \hat V^{-1}_\alpha\hat\pi^{I}_\alpha(t)] 
    -\lambda V^{-1}_\alpha 
    \int_{{\cal U}_\alpha} d^3\! \bm{x} \, \partial_\sbx\cdot\left(\bx\left(\hat\pi^I-\hat\pi^I_\alpha \right)\right)
    \simeq 0\,.
   \label{Eq:commutator_dilatation2}
\end{align}
The second term in the parentheses in the latter equation is 0, but it is kept to make it clear that the expression is written in terms of the hard modes, in the same manner as in the former equation. 
Here, we neglect the boundary terms, which represent contamination from the hard modes due to the variation of the physical volume of ${\cal U}_\alpha$ specified by the comoving coordinates. If we consider, {\it e.g.}, a sphere determined by the physical distance from a given point, this additional contribution will be eliminated, but here we simply assume that one can neglect these boundary terms. 
Then, we obtain 
\begin{align}
  [i \hat Q,\,  \hat \bphi{}_\alpha^{\rm H}(t,\, \bm{x})]  \simeq 
    - \lambda\, \Leex
    \hat \bphi{}^{\rm H}_\alpha (t,\, \bm{x}) \,,\qquad
  [i \hat Q,\,  (e^{-3\hat\zeta} \hat \bpi_\alpha)^{\rm H}(t,\, \bm{x})] \simeq 
    - \lambda\, \Leex
     \left( (e^{-3\hat\zeta} \hat \bpi)^{\rm H}_\alpha (t,\, \bm{x})\right)\,. 
    \label{Eq:commutator_dilatation_zetaH}
\end{align}

Now, we define the normalized eigenstates of the operators 
\begin{align}
   \displaystyle \hat a^I_\alpha \equiv \frac1{\sqrt{2}} \left(\frac{\hat\phi^I_\alpha}{\sigma_I}+i{e^{3\rho}} L^3 \sigma_I\hat\pi_{I\alpha}\right) \,, 
   \label{def:hata}
\end{align}
where $\sigma_I$ is an arbitrary parameter having the dimension of $\hat\phi^I$ and the summation over $I$ is not taken\footnote{Here, for our purpose, we choose the width parameter $\sigma^I$ so that the wave packet in the direction of squeezing of the wavefunction is sufficiently narrow for massless or light fields. In particular, we stress that $\sigma^I$ should be regarded as a resolution parameter of the coherent-state decomposition, rather than as a physical variance of the soft field. One may improve the choice of coherent states by adopting a linear combination of the current creation and annihilation operators as another annihilation operator. 
}. 
Using $\hat a^I_\alpha$, we define the coherent states of the soft modes by 
\begin{align}
  \hat a^I_\alpha \ket{\beta^I_\alpha} = \beta^I_\alpha \ket{\beta^I_\alpha} \,, 
\label{Def:eigenstate}
\end{align}
and $$\displaystyle \ket{\vec\bbeta}=\prod_{\alpha, I}\ket{\beta^I_\alpha}.$$
Here, $\vec\bbeta$ denotes an $(N+1)\times(\text{number of }\alpha\text{'s})$ matrix, where the arrowed and bold variables represent the vector indices with respect to $\alpha$ and $I$, respectively. Although $\ket{\vec\bbeta}$ is time-dependent, we omit the argument $t$, for notational simplicity, because we use this basis only for the projection at a fixed time mostly at the final time $t_f$.  
Although $\hat a^I_\alpha$ and $\hat a^{I \dagger}_\alpha$ are defined in the same way as the linear theory, we should note that $\phi^I$ and $\pi_I$ are the Heisenberg operators. 

Using 
\begin{align}
   [i \hat Q,\,   \hat a^I_\alpha]  \simeq - \frac{\lambda}{\sqrt{2} \sigma_I} \delta^I_0 \,,
\end{align}
which can be derived from Eq.~(\ref{Eq:commutator_dilatation_zetaH}), one obtains\footnote{
If we parametrize $\beta^I_\alpha$ as
\begin{align}
    \beta^I_\alpha \equiv\frac1{\sqrt{2}} \left(\frac{\phi^I_\alpha}{\sigma_I}+i{e^{3\rho}} L^3\sigma_I \pi_{I\alpha}\right)\,, 
\end{align}
the differentiation
$\displaystyle \lambda_\alpha \frac{\partial}{\partial {\rm Re}[\beta_\alpha^0]}$
in Eq.~\eqref{Eq:condition_dilatationinvstate_soft} can be simply written as 
$\displaystyle \lambda \frac{\partial}{\partial \zeta_\alpha}$. 
} 
\begin{align}
  i \hat Q \ket{\bbeta_\alpha} \simeq \lambda_\beta \frac{\partial}{\partial  {\rm 
  Re} [\beta^0_\alpha ] }\ket{\bbeta_\alpha} \,,  \label{Eq:condition_dilatationinvstate_soft}
\end{align}
with 
\begin{align}
    \lambda_\beta \equiv\lambda/(\sqrt{2} \sigma_0)\,.  \label{Eq:lambda}
\end{align}
This means that, when the coherent label $\beta^0_\alpha$ is real, the coherent state $|\beta^0_\alpha\rangle$ can be constructed by operating the soft part of $\hat{Q}$ on the vacuum state defined by $\hat a^I_\alpha$\footnote{The soft charge which satisfies Eq.~(\ref{Eq:commutator_dilatation2}) can be constructed as $\hat{Q}_{\rm S} = \sum_\alpha \hat{Q}_{{\rm S},\, \alpha}$ with
\begin{align}
     \hat{Q}_{{\rm S},\, \alpha}[\lambda_\beta] \equiv- \lambda e^{3\rho} L^3  \hat{\pi}_{0\, \alpha} =  i \lambda_\beta  \left( \hat{a}^0_\alpha - \hat{a}^{0 \dagger}_\alpha \right)
\end{align}
An operation of the soft charge 
\begin{align}
  e^{i\hat{Q}_{{\rm S},\alpha}[\lambda_\beta]} = e^{\lambda_\beta (\hat{a}^{\,0\dagger}_{\alpha} - \hat{a}^{\,0}_{\alpha}) }   \label{Exp:softoperation}
\end{align}
corresponds to operating the displacement operator $D(\alpha) =\exp (\alpha \hat{a}^\dagger - \alpha^* \hat{a})$ with the coherent state parameter $\alpha$ being the real number $\alpha = \alpha^* = \lambda_\beta$. Therefore, an operation of $\exp [i\hat{Q}_{{\rm S},\, \alpha}[\lambda_\beta]]$ displaces the coherent state $|\beta_\alpha^0\rangle$ by the real parameter $\lambda_\beta$ as $\left|\beta_\alpha^0 + \lambda_\beta\right\rangle
=e^{i\hat{Q}_{{\rm S}, \alpha}[\lambda_\beta]} |\beta_\alpha^0\rangle$. Expanding this expression, one obtains
\begin{align}
 i \hat{Q} |\beta_\alpha^0\rangle =    i\hat{Q}_{{\rm S},\, \alpha} [\lambda_\beta] |\beta_\alpha^0\rangle
= \lambda_\beta 
\frac{\partial}{\partial {\rm Re}[\beta_\alpha^0]}
|\beta_\alpha^0\rangle.
\end{align}
One can also confirm this by using the concrete form, (\ref{Exp:softoperation}). 
}.

\subsection{Locality condition}
Let us decompose the wave function of the universe $| \Psi \rangle$ as
\begin{align}
  \ket{\Psi} =  \int {\cal D}\bphiScvec\, \psi(\bphiScvec)\, \ket{\bphiScvec} \psiH     \,,
  \label{eq:generalWF}
\end{align}
where $\psi(\bphiScvec)$ and $\psiH$ are defined by the relation 
\begin{align}
    \psi(\bphiScvec)\, \psiH \equiv \left\langle\bphiScvec\ket{\Psi}\right.,
\end{align}
and the state for hard modes $\psiH$ is properly normalized by factoring out the wave function weight, while the wave function of the long-wavelength modes $\psi(\bphiScvec)$ is chosen to be real by absorbing the phase into $ \psiH $. The second argument $\bphiScvec$ of $\psiH$ represents that the state for hard modes $ \psiH $ depends on the set of the eigenvalues $\bphiScvec$. The integration measure for each pair of $\alpha$ and $I$ is defined as 
\begin{align}
   \int  D \beta^I_\alpha  \equiv\frac{1}{\pi}
   \int^{\infty}_{- \infty} d {\rm Re}[\beta^I_\alpha] \int^{\infty}_{- \infty} d {\rm Im}[\beta^I_\alpha] \,
\end{align}
with which the entire integration measure is given by $\displaystyle \int {\cal D}\bphiScvec \equiv\prod_{\alpha,I} \int  D \beta^I_\alpha$. 

\begin{figure}[t]
    \centering
    \includegraphics[width=0.7\linewidth,trim={0 5cm 0 0},clip]{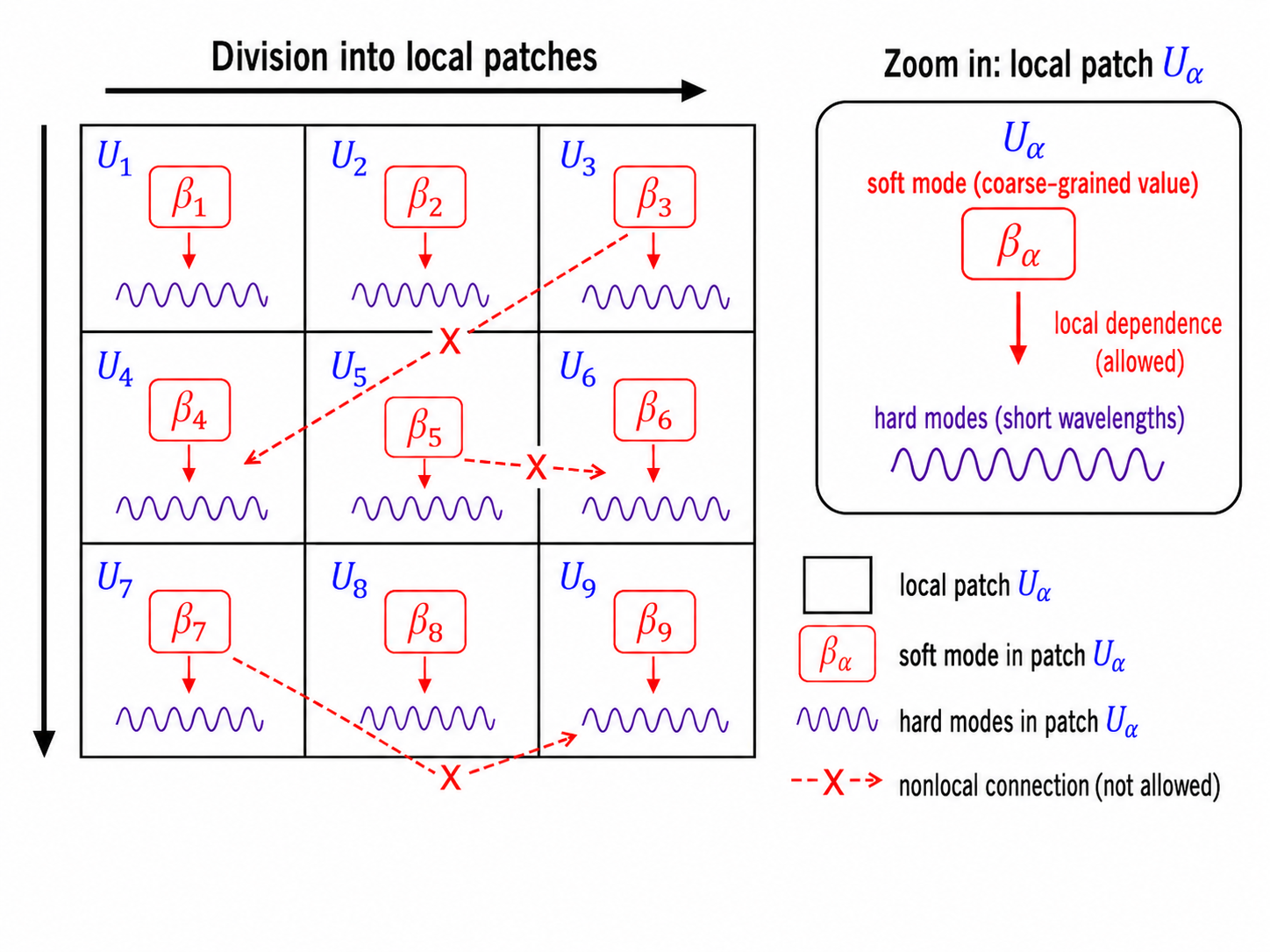}
    \caption{
    Schematic illustration of the locality condition.
    After dividing the spatial slice into local patches $U_\alpha$, the coarse-grained soft modes in each patch are labeled by $\bbeta_\alpha$.
    The locality condition requires that the hard-mode state in $U_\alpha$ depends only on the local soft-mode variables $\bbeta_\alpha$, leading to the factorized form, given in Eq.~(\ref{eq:locality}). Dashed lines represent non-local dependence on distant soft modes, which is not allowed.
    }
    \label{fig:locality_condition}
\end{figure}

Following Ref.~\cite{Tanaka:2017nff}, we require the following two conditions. First, we impose the dilatation invariance of the quantum state of the whole universe, {\it i.e.},
\begin{align}
    \hat Q\ket{\Psi}=0  \,. \label{Cond:dilatation}
\end{align}
Second, we impose a condition
\begin{align}
 \psiH=\prod_\alpha \psiHlocal,
 \label{eq:locality}
\end{align}
where $\psiHlocal$ represents the hard-mode state in each local universe ${\cal U}_\alpha$. 
This condition indicates that the hard-mode state depends on the soft modes only through the local values in the same local patch ${\cal U}_\alpha$. For this reason, we call this the locality condition. When this condition holds, after coarse-graining the hard modes, the effective dynamics of soft modes (or more specifically, super-Hubble modes) can be solved using the gradient expansion. Although the influence of soft modes in the local universe ${\cal U}_\alpha$ can be labeled solely by $\bbeta_\alpha$, this condition does not prohibit a possible correlation among soft modes in different patches, $\bbeta_\alpha$ and $\bbeta_{\alpha'}$ with $\alpha \neq \alpha'$. 
The long-range correlations can be encoded as the non-separability of the wave function of the soft modes, {\it i.e.}, $\psi(\bphiScvec) \neq \prod_\alpha \psi_\alpha (\bbeta_\alpha)$. 

In Ref.~\cite{Tanaka:2017nff}, we used the eigenstates of $\hat{\vec\zeta}$. However, if we specify the global distribution of $\vec\zeta$, we can specify $\vec\pi_\zeta$ when they are mutually strongly correlated. Hence, the local hard-mode states in ${\cal U}_\alpha$ can depend on the values of $\vec\zeta$ in the other patches through the correlation to $\pi_{\zeta\alpha}$. 
To avoid this spurious non-locality, we instead use an expansion in terms of coherent states.

Under condition (\ref{eq:locality}), the dilatation invariance of the wave function leads to  
\begin{align}
   0=\bra{\vec{\bbeta}} i\hat Q\ket{\Psi} =\psi\left(\bphiScvec\right)
\sum_{\alpha}  \left(i\hat Q-\lambda_\beta \frac{\partial}{\partial {\rm Re}[\beta^0_\alpha]}\right) \psiHlocal \prod_{\alpha' \neq \alpha }  \ket{\Psi_{\alpha'};  \bbeta_{\alpha'} } 
      -\lambda_\beta \psiH\left( \sum_\alpha \frac{\partial\psi(\bphiScvec)}{\partial {\rm Re}[\beta^0_\alpha] }\right)\,. 
\end{align}
Taking the real and imaginary parts, we obtain
\begin{align}
& \sum_\alpha \frac{\partial\psi(\bphiScvec)}{\partial {\rm Re}[\beta^0_\alpha]}=0,
\label{eq:homoWF}\\ 
&\left( i\hat Q - \lambda_\beta \frac{\partial}{\partial {\rm Re}[\beta^0_\alpha]} + i c_\alpha \right)\psiHlocal= 0\, , 
    \label{eq:localityOld}
\end{align}
where $c_\alpha$ is real and satisfies $\displaystyle\sum_\alpha c_\alpha =0$. In what follows, we omit the third term of Eq.~(\ref{eq:localityOld}), attributing the appropriate phase factor to the respective $\psiHlocal$. The condition \eqref{eq:homoWF}, which can be rewritten as $\partial \psi(\bphiScvec)/\partial \bar\zeta = 0$ with $\bar{\zeta}$ being the spatial average of $\zeta$ over the entire spacelike region, requires the homogeneity of the soft-mode wave function in the direction of $\bar\zeta$. This condition must be satisfied to avoid violating the dilatation invariance. In this way, starting from \eqref{eq:locality}, one can also derive a variant of the locality condition (\ref{eq:localityOld}), introduced in Ref.~\cite{Tanaka:2017nff}. 

\subsection{Weinberg's adiabatic mode}  \label{SSec:WAM}
The time-independent solution for $\zeta$ in the soft limit, known as Weinberg's adiabatic mode~\cite{Weinberg:2003sw}, has been discussed in relation to large gauge transformations and cosmological soft theorems~\cite{Hinterbichler:2012nm, Jazayeri:2017huf, Tanaka:2017nff, Mitsou:2021aue}. In this subsection, we show that Weinberg's adiabatic mode generically exists as an operator solution when the locality condition (\ref{eq:locality}) holds. This generalizes our discussion in Refs.~\cite{Tanaka:2015aza, Tanaka:2017nff}, which showed the existence of Weinberg's adiabatic mode as a solution of the effective action. For this purpose, we introduce a dilatation operator as 
\begin{align}
  \hat {\tilde Q}=\int d^3\! \bx\, W(\bx) \hat q(t,\,\bx)\,, \label{Def:Q0}
\end{align}
using $\hat q$ introduced in Eq.~\eqref{Eq:charge_dilatation}. 
Here, $W(\bx)$ is a smooth window function which remains almost constant in each separate universe ${\cal U}_\alpha$. When we set the window function to 1, $\hat {\tilde Q}$ becomes $\hat {Q}$.  Therefore, one can approximate $\hat {\tilde Q}$ by $W(\bx_\alpha) \hat Q $ when it acts on the quantum state in $\cal{U}_\alpha$. 

Operating $ \exp(-i \hat{\tilde Q})$ on a quantum state $| \Psi \rangle$ in Eq.~\eqref{eq:generalWF},  
we obtain
\begin{align}
\ket{\Psi'}
  &=
     \int {\cal D}\bphiScvec\,
      \psi(\bphiScvec)
      \exp\left(-i\hat{\tilde Q}\right) \ket{\bphiScvec}
      \left[\prod_\alpha \psiHlocal\right]\cr
  & \simeq \int {\cal D}\bphiScvec\, 
      \psi(\bphiScvec)
        \prod_{\gamma} \left[ \exp\left(-\lambda_\beta  W(\bx_\gamma)\frac{\partial}{\partial {\rm Re}[\beta^0_\gamma]}\right)
      \ket{\bbeta_\gamma} \right]
      \left[\prod_\alpha \psiHlocal\right]\cr   
  &=\int {\cal D}\bphiScvec\,
      \psi(\bphiScvec_\lambda) \ket{\bphiScvec}
      \left[\prod_\alpha \psiHlocal\right]\,,
  \label{eq:WAM_wave function}
\end{align}
where $\bphiScvec_\lambda$ is defined as $\bphiScvec_\lambda \equiv( \vec{\beta}^0_\lambda,\, \vec\bbeta' )$, where $\vec{\bbeta}_\alpha'$ only counts $I=1,\, 2,\, \cdots$, with $\vec{\beta}^0_\lambda$ being shifted by the real $\gamma$-dependent parameter as 
\begin{align}
    \beta^0_{\gamma, \lambda} \equiv
    \beta_\gamma^0 + \lambda_\beta W(\bx_\gamma)\,.
\end{align}
This simply corresponds to a time-independent shift of the value of $\zeta_\alpha \equiv \sqrt{2}\sigma_0 {\rm Re}[\beta^0_\alpha]$ by $\lambda$ in the long wavelength perturbations of $\zeta$, since $\lambda_\beta$ is related to $\lambda$ via Eq.~(\ref{Eq:lambda}).

Precisely speaking, the change from $| \Psi \rangle$ to $| \Psi' \rangle$ does not correspond to a solution, because the commutation relation between the Hamiltonian 
\begin{align}
 \hat H=\int d^3\! \bx\, e^{3(\rho+\hat\zeta)} \hat{\cal H}(\bphi,e^{- 3\hat\zeta}\hat{\bm{\pi}},e^{-\hat\zeta}\partial_i, \Delta^{-1})\,,
\end{align}
and $\hat{\tilde Q}$ is given by 
\begin{align}
 [i\hat {\tilde Q},\hat H]=\int d^3\!\bx\, (\bx\cdot\partial_{\sbx} W(\bx)) e^{3\hat\zeta}\hat{\cal H}(t,\bx)\,, 
 \label{eq:commutationQH}
\end{align}
which does not vanish. Here, we used the fact that $\hat{\cal H}$ transforms as a scalar under the dilatation transformation. However, since the residual in \eqref{eq:commutationQH} is suppressed for a sufficiently smooth window function $W$, at the leading order of the gradient expansion, one can recognize $| \Psi' \rangle$ as a proxy of the solution, corresponding to Weinberg's adiabatic mode. One would be able to modify $\hat{\tilde{Q}}$ at the sub-leading order of the gradient expansion so that the non-zero contribution in Eq.~(\ref{eq:commutationQH}) is canceled. As such, one can construct an operator that resembles $\hat{\tilde Q}$ but commutes with $\hat H$ at each order of the gradient expansion. 

\subsection{Dilatation-invariant operators}
In Refs.~\cite{Urakawa:2010it, Urakawa:2010kr}, genuine gauge-invariant operators were introduced as operators insensitive to the variation of gauge conditions outside the observable regions. As a special class, these residual gauge degrees of freedom include large gauge transformations such as the dilatation~(see Refs.~\cite{Hinterbichler:2012nm, Tanaka:2013caa} for a systematic argument on large gauge transformations in cosmology). 

In this paper, among various infrared pathologies, {\it e.g.}, summarized in Ref.~\cite{Tanaka:2013caa}, we focus on the IR divergences that arise from the momentum integrations of Wightman propagators. 
For this purpose, it is sufficient to concentrate on dilatation-invariant operators\footnote{ 
Genuine gauge-invariant operators, whose concrete construction was introduced in Refs.~\cite{Urakawa:2010it, Urakawa:2010kr}, are necessary to ensure the suppression of all infrared enhancements, including the secular growth~(see e.g., \cite{Weinberg:2005vy}). In Refs.~\cite{Urakawa:2010it, Urakawa:2010kr, Tanaka:2011aj, Tanaka:2013caa, Tanaka:2013xe, Tanaka:2014ina}, the infrared regularity of the genuine gauge-invariant operators was shown for a specific quantum state, {\it i.e.}, the Euclidean vacuum, providing a complementary explanation to the discussion based on the gauge fixing~\cite{Urakawa:2009my, Urakawa:2009gb}.}. 
We introduce a simplified version of genuine gauge-invariant operators, which are invariant only under the dilatation transformation, satisfying 
\begin{align}
 [i\hat Q,\giop ]=0\,.
\end{align}
The so-called ordinary gauge-invariant variables, {\it e.g.}, $\zeta$, are not invariant under dilatation transformations.
When $\giop$ is a local operator composed of $\bphiS$, $\bpi_\alpha(t)$ and the hard-mode fields in ${\cal U}_\alpha$, we refer to it as a local dilatation-invariant operator, which is denoted by $\lgop$. For example, a dilatation-invariant operator corresponding to a local operator $\hat{\cal O}_\alpha$, which is invariant under small-gauge transformation, can be constructed by specifying the spatial coordinates in terms of the geodesic coordinates $\bX$. As an approximation, we consider the geodesic coordinates neglecting all metric perturbations except for $\wam$, {\it i.e.}, $\bX\equiv e^\wam \bx$. 
For notational brevity, we introduce the spacetime coordinates as $X \equiv(t,\, \bX )$. The dilatation-invariant counterpart of $\hat{\cal O}_{\alpha}$ is given by 
\begin{align}
  {}^{\raisebox{0.4ex}{{\scriptsize$g$}}}\!\hat{\cal O}_{\alpha} \left(\{X_i\}\right)
  =\hat{\cal O}_{\alpha}\left(\{(t_i,\, e^{-\hat{\zeta}_\alpha (t_i)} \bX_i)\}\right)
  =\biggl(1- \sum_j \hat{\zeta}_\alpha (t_j) \bX_j\cdot\frac{\partial}{\partial \bX_j} 
   +\cdots\biggr) \hat{\cal O}_{\alpha}\left(\{X_i\}\right)\,,
  \label{eq:giexp}
\end{align}
where the set of arguments specifies the coordinate values of the fields contained in the operator.

We decompose $\wam$ into $\hat{\zeta}_{\alpha \star} \equiv\hat{\zeta}_{\alpha}(t_\star)$ and corrections as
\begin{align}
  \wam=\hat{\zeta}_{\alpha \star} +\delta\wam.  \label{Eq:zetadecomposition} 
\end{align}
We assume that the comoving scale $L$ is well beyond the horizon scale at $t=t_\star$, {\it i.e.}, $L \gg (e^{\rho_\star} \dot{\rho}_\star)^{-1}$. Since $\hat Q$ is time-independent, the dilatation transformation of $\hat{\zeta}_{\alpha \star}$ is given by 
\begin{align}
 [i\hat Q,\hat{\zeta}_{\alpha \star}]=-\lambda\,. 
\end{align}
Combined with Eq.~\eqref{Eq:commutator_dilatation2}, we obtain 
\begin{align}
[i\hat Q,\delta\wam]=0\,, 
\label{eq:GIdeltazeta}
\end{align}
which implies that $\delta\wam$ is a dilatation-invariant operator. 
In the ordinary models which do not allow superluminal propagation of information, the superhorizon evolution should be local. Hence, we conclude that $\delta\wam$ should be a {\it local dilatation-invariant} operator.

\section{Generalization of Soft Theorem}  \label{Sec:ST}
In this section, we show that when the locality condition (\ref{eq:locality}) holds, one can always factor out the correlators for the soft modes. It will turn out that this factorization formula reduces to standard soft theorems in the context of inflation, which is also known as consistency relations~\cite{Creminelli:2004yq,Weinberg:2003sw}, when two additional conditions hold. 

For this purpose, we begin with the approximation
\begin{align}
\bra{\Psi} f\!\left(\hat{\vec\zeta}\right)\,{}^g\!\hat{\cal O}_\alpha\ket{\Psi}
\simeq 
 \int{\cal D}\vec\bbeta\, 
  \psi^2(\vec\bbeta) 
  \bra{\vec\bbeta}
  f\!\left(\hat{\vec\zeta}\right)\, \psiHlocalbra \lgop \psiHlocal
  \ket{\vec{\bbeta}},  
  \label{eq:KeyApprox}
\end{align}
where $f\!\left(\,\wamvec \,\right) $ is an arbitrary function of $\wamvec$.
The meaning of ``$\simeq$'' is that the ambiguity from the operator ordering of soft modes is neglected. 
In other words, it is assumed that the extension of the wave packet corresponding to each coherent state $\ket{\vec\bbeta}$ is negligibly small. 
This formula tells us that we can identify the coherent state parameters $\vec\bbeta$ and $\vec{\tilde\bbeta}$ in the bra and ket vectors, up to these errors. 
The validity of this approximate equality is discussed in Appendix \ref{AppB}.
Here and hereafter, for simplicity, we only consider $\lgop$ which consists only of operators at $t=t_f$\footnote{When $\lgop$ contains operators at different times, {\it e.g.}, at $t_f$ and $t$, there appears a unitary operator for the time evolution between $t$ and $t_f$. Since the Hamiltonian is also dilatation invariant, a similar argument also applies. }.

Introducing 
{\begin{align}
\bre{\hat X}_{\!\sbm{\beta}_\alpha}
\equiv \bra{\bbeta_\alpha} \psiHlocalbra \hat{X} \psiHlocal \ket{\bbeta_\alpha},   \label{Def:Xcoh}
\end{align}
for an arbitrary operator $\hat{X}$, we first show that $\lgopbre$ satisfies
\begin{align}
    \frac{\partial}{\partial \zeta_\alpha} \lgopbre = 0\,. \label{Eq:locality2}
\end{align}
{Equation (\ref{Eq:locality2}) can be shown from a more general formula
\begin{align}
    0 = \bra{\bbeta_\alpha}\psiHlocalbra   \left[\lgop,\, i \hat{Q} \right] \psiHlocaltilde \ket{\tilde\bbeta_\alpha}
   & = \lambda 
     \bra{\bbeta_\alpha}\psiHlocalbra  \left( \frac{\overleftarrow{\partial}}{\partial \zeta_\alpha} \lgop + \lgop \frac{\partial}{\partial \tilde\zeta_\alpha} \right)  \psiHlocaltilde \ket{\tilde\bbeta_\alpha}\cr 
   & = \lambda \frac{\partial}{\partial \zeta^c_\alpha} 
  \bra{\bbeta_\alpha} \psiHlocalbra \lgop \psiHlocaltilde   \ket{\tilde\bbeta_\alpha}\,, 
 \label{Eq:derivation_locality2}
\end{align}
where $\tilde\zeta_\alpha=\sqrt{2} \sigma_0 {\rm Re} \tilde\beta^0_\alpha$ and 
the partial differentiation with respect to $\zeta^c_\alpha \equiv (\zeta_\alpha+\tilde\zeta_\alpha)/2$ is taken for a fixed value of $\zeta^\Delta_\alpha \equiv (\zeta_\alpha-\tilde\zeta_\alpha)/2$. 
Therefore, using Eq.~(\ref{Eq:locality2}) and introducing the non-adiabatic components $\bbeta_{{\rm nad}, \alpha}$, defined as
\begin{align}
   \bbeta_{{\rm nad}, \alpha} = \pi_{\zeta\alpha}~ {\rm and} \,\,\beta^I_\alpha\,\, (I=1,\, 2,\, \cdots ) \,,
\end{align}
$\lgopbre$ can be expanded as
\begin{align}
    \lgopbre = \lgopbre\Bigr|_{\sbbeta_{\rm{nad}, \alpha}=0} + \frac{\partial \lgopbre}{\partial \beta{}^a_{{\rm nad}, \alpha}} \Bigg|_{\sbtheta_{, \alpha}=0}\!\! \beta{}^a_{{\rm nad}, \alpha}  + \cdots, \label{Exp:gOalpha}
\end{align}
where $a$ labels the field index of $\bbeta_{{\rm nad}, \alpha}$.}

Taking into account that the use of the soft-mode wave function, or equivalently the quantum state of long-wavelength modes, has also played an important role in derivations and discussions of inflationary soft theorems~\cite{Goldberger:2013rsa, Pimentel:2013gza,Garriga:2016poh,Hui:2018cag}, we divide the following discussion into two cases according to the structure of the soft-mode wave function. 

\begin{enumerate}
    \item {\bf Separable case} 
    
When the wave function $\psi(\vec\bbeta)$ is approximately separable for the adiabatic perturbation from the other soft variables,  
\begin{align}
\psi\left(\vec\bbeta\right) = \psi_\zeta\left(\vec{\zeta}  \right) \times \psi_{\rm nad}\left(\vec{\bbeta}_{\rm nad} \right)\,,
\label{eq:factorization_psi}
\end{align}
we obtain 
\begin{align}
 \bbra{\Psi}f\!\left(\,\wamvec\, \right) \lgop\bket{\Psi}
 \simeq \frac{\int D \vec{\zeta}\, \psi^2_\zeta (\vec{\zeta})  
    f\!\left(\,\vec\zeta\,\right)}{\int {\cal D} \vec{\bbeta}\, \psi^2 (\vec{\bbeta})}  
 \times \left[\int {\cal D} \vec{\bbeta}_{\rm nad} \, 
    \psi^2_{\rm nad}(\vec{\bbeta}_{\rm nad} )\lgopbre\right]
    =\bbra{\Psi}f\!\left(\,\wamvec\, \right) \bket{\Psi}
    \times \bbra{\Psi} \lgop\bket{\Psi} \,,  
    \label{Exp:decomposition_separable} 
\end{align}
where we introduce a normalized expectation value
\begin{align}
\bbra{\Psi} \hat O \bket{\Psi}\equiv
\frac{\bra{\Psi}\hat O  \ket{\Psi}}{\bra{\Psi}\Psi\Bigr\rangle}\,,
\label{normexp}
\end{align}
for an arbitrary operator $\hat O$, and ${\cal D} \vec\bbeta_{\rm nad}$ so as to satisfy ${\cal D} \vec{\bbeta} = D \vec{\zeta}\, {\cal D} \vec\bbeta_{\rm nad}$. In Eq.~(\ref{Exp:decomposition_separable}), the first factor on the right-hand side corresponds to the correlators of $\hat\zeta$, while the second one corresponds to those for the other modes including hard modes. 
The denominator $ \langle \Psi|\Psi \rangle$ in Eq.~\eqref{normexp} is necessary, because the wave function $| \Psi \rangle$ is not normalizable, owing to the existence of the flat direction corresponding to $\bar\zeta$. 
Equation~\eqref{eq:factorization_psi} states two conditions: one is the suppression of the conjugate momentum $\hat{\pi}_\zeta$ and the other is the absence of correlation between $\hat{\zeta}$ and the soft isocurvature modes. 

In what follows, let us show that Eq.~\eqref{Exp:decomposition_separable} corresponds to the so-called consistency relation which was first derived by Maldacena for superhorizon modes \cite{Maldacena:2002vr}. For this purpose, let us introduce the soft mode operators in Fourier space as 
\begin{align}
  \hat{\cal O}{}_{\sbk}
   \equiv\frac1{(2\pi)^3}\sum_\alpha e^{- i \sbk\cdot\sbx_\alpha} 
   \hat{\cal O}_\alpha\,. 
\end{align}
We use $\bk$ to specify the momentum of soft modes, while $\bp$ to specify those of hard modes contained in $\hat{\cal O}^{\rm H}$. 
With this definition, $\wamk \simeq 0$ holds for $k>\pi L^{-1}$. Choosing $f\! \left(\,\wamvec\,\right) = \wamvec$ and rewriting the soft indices into the Fourier modes, one obtains
\begin{align}
 \bbra{\Psi} \wamk\, {}^g\hat{\cal O}_{\sbk'}\bket{\Psi}
  =\bbra{\Psi} \wamk \bket{\Psi}\times  \bbra{\Psi} {}^g\hat{\cal O}_{\sbk'}\bket{\Psi} =0\,, 
  \label{eq:vanishingcorrelation}
\end{align}
because $\bbra{\Psi} \wamk \bket{\Psi}=0$. 
Substituting Eq.~\eqref{eq:giexp} into the left-hand side of this relation, and using the relation between an arbitrary function $g(\bx)$ and its Fourier transform $\tilde g(\bp)$,
\begin{align}
  \int d^3\!\bx\, e^{-i\sbm{p}\cdot\sbm{x}}  \bx\cdot \frac{\partial}{\partial \bx} g(\bx)
 =-\frac{\partial}{\partial \bp} \cdot \bp ~\tilde g(\bp)\,,
\end{align}
one can derive the consistency relation as 
\begin{align}
 \bbra{\Psi}\wamk \hat{\cal O}_{\sbk'}(\{\bp_i\})\bket{\Psi}'
  \simeq -\bbra{\Psi}\wamk \wammk \bket{\Psi}' 
   \left(\sum_i \frac{\partial}{\partial \bp_i}\cdot \bp_i \right)
    \bbra{\Psi} \hat{\cal O}_{-\sbk-\sbk'}(\{\bp_i\})\bket{\Psi}'\,,
 \label{eq:standardCR}
\end{align}
where ``\,$\bbra{\Psi}\cdots \bket{\Psi}'$\,'' denotes a correlator after removing the delta function $\delta(\bk+\bk')$. 
In this way, the well-known consistency relation~\cite{Maldacena:2002vr, Creminelli:2004yq, Pimentel:2012tw} can be reproduced from Eq.~\eqref{eq:locality} under the separability condition of the wave function, (\ref{eq:factorization_psi}).  

\item {\bf Non-separable case} \\
In general, the wave function does not take the separable form of \eqref{eq:factorization_psi}. 
Then, the corresponding soft theorem deviates from the consistency relation. The condition (\ref{eq:factorization_psi}) can be violated in the presence of multiple light fields or when the conjugate momentum of $\zeta$ is not negligibly small, {\it i.e.}, when $\zeta$ exhibits significant time variation in the soft limit. A typical example of the latter is ultra slow-roll inflation~\cite{Kinney:2005vj}, or more generally a non-attractor single-field phase, in which the conjugate momentum of $\zeta$ is not negligible and the standard single-field consistency relation is modified~\cite{Namjoo:2012aa, Martin:2012pe, Chen:2013aj, Chen:2013eea, Suyama:2021adn}.

Using the expansion~\eqref{Exp:gOalpha}, we have 
\begin{align}
  \bbra{\Psi}f\!\left(\,\wamvec\, \right) \lgop\bket{\Psi}
 \simeq & 
\bbra{\Psi} f\!\left(\,\wamvec\, \right)\bket{\Psi}
     \left[\lgopbre\right]_{\sbtheta_{, \alpha}=0} 
     + \bbra{\Psi}f\!\left(\,\hat{\vec\zeta}\,\right)\hat\beta{}^a_{{\rm nad}, \alpha}\bket{\Psi} 
     \left[\frac{\partial \lgopbre}{\partial \beta{}^a_{{\rm nad}, \alpha}}\right]_{\sbtheta_{,\alpha}=0} +\cdots
 \,,  
\label{eq:psi_pertmixing}
\end{align}
where we used the relation
\begin{align}
\bbra{\Psi} {\cal O}(\hat{\vec\bbeta})\bket{\Psi} \simeq 
   \frac{\int {\cal D} \vec{\bbeta}\, \psi^2 (\vec{\bbeta})  
   {\cal O}\left(\,\vec\bbeta\,\right)} 
    {\int {\cal D} \vec{\bbeta}\, \psi^2 (\vec{\bbeta})}, 
\end{align}
which holds for an arbitrary soft-mode operator ${\cal O}(\hat{\vec\bbeta})$. 

As is shown here, when the locality condition (\ref{eq:locality}) holds, one can factor out the correlation functions for the soft modes. To the best of our knowledge, Eq.~\eqref{eq:psi_pertmixing} is the most general soft theorem associated with the dilatation among those which have been derived so far. 

\end{enumerate}

\section{Perturbative suppression of loop corrections} \label{Sec:Loops}
In this section, we show that the locality condition (\ref{eq:locality}) universally constrains loop corrections to correlators of the coarse-grained curvature perturbation $\wam$, leading to their perturbative suppression when correlations with other light fields are suppressed.

\subsection{Two-point function}
For notational brevity, first let us start with the two-point function, $\bbra{\Psi} \hat{\zeta}_{\alpha}(t)  \hat{\zeta}_{\alpha'}(t) \bket{\Psi} $, which can be expanded as
\begin{align}
    \bbra{\Psi} \hat{\zeta}_{\alpha}(t_f)  \hat{\zeta}_{\alpha'}(t_f) \bket{\Psi}  
    = \bbra{\Psi} \hat{\zeta}_{\alpha \star} \hat{\zeta}_{\alpha' \star}   \bket{\Psi} + \bbra{\Psi} \hat{\zeta}_{\alpha \star} \delta \hat{\zeta}_{\alpha'}(t_f) \bket{\Psi}+ \bbra{\Psi} \delta \hat{\zeta}_{\alpha}(t_f) \hat{\zeta}_{\alpha' \star}  \bket{\Psi} + \bbra{\Psi}  \delta \hat{\zeta}_{\alpha}(t_f) \delta \hat{\zeta}_{\alpha'}(t_f) \bket{\Psi}\,, \label{Eq:expansion2pt}
\end{align}
by using Eq.~(\ref{Eq:zetadecomposition}). In what follows, we show that the loop corrections, described by the last three terms, are perturbatively suppressed compared to the linear power spectrum, described by the first term, apart from the standard superhorizon evolution sourced by isocurvature perturbations (see e.g., \cite{Gordon}). This argument generalizes Ref.~\cite{Tanaka:2015aza}, where it was shown that radiative corrections from heavy fields do not significantly modify the superhorizon evolution of the curvature perturbation. As discussed in Eq.~\eqref{eq:GIdeltazeta}, the non-linear part of $\wam$, {\it i.e.}, $\delta\wam$ is dilatation invariant. 
Since $\delta\wam$ corresponds to $\lgop$, one can repeat a similar argument to the previous section
to show the suppression of the loop corrections in the presence of the non-adiabatic soft modes. 

Using Eq.~(\ref{eq:psi_pertmixing}) with $f\!\left(\,\wamvec\,\right) = \hat{\zeta}_{\alpha \star}$ and $\lgop= \delta\wam$, the second term can be rewritten as 
\begin{align}
\bbra{\Psi} \hat\zeta_{\alpha\star}\delta\hat\zeta_{\alpha'}(t_f)\bket{\Psi} 
 \simeq
 \bbra{\Psi}\hat\zeta_{\alpha\star}\hat\pi{}_{\zeta,\alpha'}\bket{\Psi} 
    \! \left[ \frac{\partial \bre{\delta \hat{\zeta}_{\alpha'}}_{\sbm{\beta}_{\alpha'}}}{\partial \pi{}_{\zeta,\alpha'}} 
    \right]_{{\sbbeta}_{\rm nad,\alpha'}={\sbm 0}} \hspace{-10pt}
    + \,\,\,\,\sum_{I=1}^N \bbra{\Psi}\hat\zeta_{\alpha\star}\hat\beta^I_{\alpha'}\bket{\Psi} 
    \left[ \frac{\partial \bre{\delta \hat{\zeta}_{\alpha'}}_{\sbm{\beta}_{\alpha'}}
    }{\partial \beta^I_{\alpha'}}
    \right]_{{\sbbeta}_{\rm nad,\alpha'}={\sbm 0}}\hspace{-15pt} +\cdots\,, \label{Eq:loop2nd}
\end{align}
where $\langle \delta \hat{\zeta}_{\alpha'} \rangle_{\sbm{\beta}_{\alpha'}}$ is defined by Eq.~(\ref{Def:Xcoh}) with $X$ being $\delta \hat{\zeta}_{\alpha'}$. These two terms are both part of the expansion with respect to $\bbeta_{{\rm nad},\alpha'}$, but we have written them separately for the sake of the following explanation. The first term characterizes the contribution associated with a non-negligible conjugate momentum of the curvature perturbation, $\hat{\pi}_\zeta$, as in a non-attractor or non-slow-roll phase. During a non-attractor epoch, a non-negligible $\hat{\pi}_\zeta$ can source the time evolution of $\zeta$ even after horizon crossing. For moderately soft modes, the curvature perturbation can be generated after horizon crossing through $O(k^2)$ terms in the gradient expansion. Although these corrections are small at each instant, they can accumulate and give a sizable contribution~\cite{Leach:2001zf}. The generated curvature perturbation can remain after the end of the non-attractor epoch, but the magnitude of the conjugate momentum $\pi_\zeta$ has the wavenumber dependence $\propto k^2$ for the soft modes that are already beyond the horizon scale before the non-attractor epoch starts. Hence, softer modes are, in general, more suppressed in this regime. Meanwhile, the second term characterizes the contribution from isocurvature perturbations, which can also affect the superhorizon evolution of $\zeta$ without being suppressed by the soft wavenumber $k$. As has been widely investigated, both terms can be important already at linear order when the non-attractor epoch or the multi-field epoch lasts sufficiently long.

In addition to such linear contributions, the terms in Eq.~(\ref{Eq:loop2nd}) can also generate loop corrections through the hard-mode dependence contained in $\delta \hat{\zeta}_{\alpha}$. Under the locality condition, such hard-mode loop corrections can enter only through correlations with the non-adiabatic soft variables, namely the soft mode of $\hat{\pi}_\zeta$ or the isocurvature perturbations, but not through an undifferentiated soft $\hat{\zeta}$, in contrast to the situation discussed in Ref.~\cite{Kristiano:2022maq}. 
In order for hard modes to leave an impact on the soft mode, 
the hard-mode sector needs to be significantly enhanced. In particular, for far-infrared modes in a single-field model of inflation, the linear mixing
$\langle\! \langle \Psi| \hat\zeta_{\alpha\star}\hat\pi{}_{\zeta,\alpha'}|\Psi \rangle \! \rangle$
is already suppressed by $k^2$. Therefore, a non-negligible loop correction on a large scale necessarily requires a breakdown of the perturbative expansion on a smaller scale.
When a transient phenomenon which enables the time-variation of $\hat{\vec{\zeta}}$ on superhorizon scales occurs, these loop corrections can be enhanced only for a moderately small $k$.

Similarly, for the last term in Eq.~\eqref{Eq:expansion2pt}, 
we obtain 
\begin{align}
   \bbra{\Psi} {\delta\hat\zeta_\alpha\delta\hat\zeta_{\alpha'}} \bket{\Psi}  
   \simeq \left\{
   \begin{array}{ll}
\displaystyle   \bre{\delta\hat\zeta_{\alpha}^2}_{\!\sbbeta_{\alpha}={\sbm 0}} 
 +   \bbra{\Psi}\hat\pi_{\zeta, \alpha}^2\bket{\Psi} 
    \left[ \frac{\partial^2 \bre{\delta\hat\zeta_{\alpha}^2}_{\sbm{\beta}_\alpha}}{\partial \pi_{\zeta,\alpha}^2} \right]_{\!{\sbbeta}_{\!{\rm nad},\alpha} = \vec{\sbm 0}}  +\cdots\,,&(\alpha=\alpha')\,,
    \label{eq:last_term}
    \cr
\displaystyle
\bbra{\Psi}\hat\pi_{\zeta, \alpha} \hat\pi_{\zeta, {\alpha'}}\bket{\Psi} 
    \left[ \frac{\partial \bre{ \delta\hat\zeta_\alpha}_{\!\sbm{\beta}_\alpha}}{\partial \pi_{\zeta,\alpha}}\right]_{\sbbeta_{{\rm nad},\alpha} = {\sbm 0}} 
    \left[ \frac{\partial \bre{ \delta\hat\zeta_{\alpha'}}_{\!\sbm{\beta}_{\alpha'}}}{\partial \pi_{\zeta,{\alpha'}}}\right]_{\sbbeta_{{\rm nad},{\alpha'}} = {\sbm 0}}+\cdots\,,&(\alpha\ne\alpha')\,.
    \end{array}\right.
\end{align}
Here, we concentrate on cases in which one can neglect the contributions from the (light) isocurvature perturbations, which can lead to an evolution of $\zeta$, unsuppressed on a large length scale. 
The first term on the right-hand side in Eq. \eqref{eq:last_term} for $\alpha=\alpha'$ is simply the contribution originating from the random fluctuation of each local patch. This kind of ultra-violet contribution to the long-wavelength spectrum takes the maximum value for the largest value of $k$, which is $\pi L^{-1}$ here, and should be suppressed as $\propto k^3$ for a small $k$. Hence, this term is also unimportant for a very long-wavelength perturbation, as long as the perturbative expansion on a small length scale remains valid. The other terms can also describe a superhorizon evolution sourced by $\hat{\pi}_\zeta$, which is suppressed in a long wavelength limit, for the same reason as in the case of $\langle\! \langle \Psi| \hat\zeta_{\alpha\star}\delta\hat\zeta_{\alpha'}(t_f) | \Psi \rangle\!\rangle$. 

\subsection{Validity of effective field theory prescription}
Up until now, we have considered the two-point function for notational brevity, but the discussion can be easily extended to higher point functions. When both the locality condition (\ref{eq:locality}) and the separability condition (\ref{eq:factorization_psi}) hold, the correlators including $\delta \hat{\zeta}$ 
are suppressed by the correlators including $\pizetavec(t_f) \simeq  \partial_t \hat{\vec{\zeta}} (t_f)$ or have only UV contributions which become negligible for $k\ll aH$. 
Thus, we obtain
\begin{align}
     \bbra{\Psi} \hat{\zeta}_{\alpha_1}(t_f)  \hat{\zeta}_{\alpha_2}(t_f) \cdots \hat{\zeta}_{\alpha_n}(t_f) \bket{\Psi}  \approx  \bbra{\Psi} \hat{\zeta}_{\alpha_1 \star} \hat{\zeta}_{\alpha_2 \star} \cdots \hat{\zeta}_{\alpha_n \star}   \bket{\Psi}\,,
\end{align}
for arbitrary $n$-point functions at an arbitrary order of perturbation in effectively single-field models of inflation. This implies that the shorter wavelength modes with $k \gg L$, which cross the horizon after $t = t_\star$, yield only suppressed corrections to the curvature perturbations at small $k \ll L$, proving the conservation of the soft modes of the curvature perturbation in the early universe. 

In this paper, we do not construct an effective description obtained by coarse-graining the hard modes.  However, since the interactions with the hard modes give rise only to contributions suppressed at small $k$, they do not generate non-local terms in the coarse-grained action that would violate the locality requirement for the separate universe approach~\cite{Tanaka:2021dww}.  In Refs.~\cite{Inomata:2024lud,Ema:2026xxx}, related cancellations of loop corrections from short-wavelength perturbations to superhorizon modes were discussed within an effective description of the soft sector, namely within the leading-order gradient expansion, assuming that this description is valid. Our argument provides a complementary perspective by clarifying the role of the locality condition in validating the effective description of soft modes. The effective dynamics of superhorizon modes has also been studied from the viewpoint of stochastic inflation and Soft de Sitter Effective Theory (SdSET), in which the stochastic evolution~\cite{Starobinsky:1986fx,Starobinsky:1994bd} emerges as the infrared effective dynamics of light fields in de Sitter space
~\cite{Cohen:2020php,Cohen:2021fzf}. (See also Refs.~\cite{Green:2024xxx} for a more recent development.)

\section{IR regularity revisited}  \label{Sec:IRreg}
As shown in Eq.~(\ref{Exp:decomposition_separable}), when we expand the correlators in terms of coherent states, the weight in calculating the expectation value $\bbra{\Psi}\cdots\bket{\Psi}$ is given by the squared wave function $\psi^2_\zeta (\vec\zeta)$, whose support extends to $\bar\zeta\to \pm\infty$.
This is essentially the origin of infrared divergences, and how they are canceled for genuine gauge-invariant correlators was already discussed in~\cite{Tanaka:2017nff}~\footnote{In this paper, we do not discuss an enhancement of the isocurvature loops in the infrared limit, which also can persist in the correlators of the genuine gauge-invariant operators. See e.g., Refs.~\cite{Urakawa:2009gb, Tanaka:2013caa} and references therein. }. Here, we provide an alternative way to understand the cancellation mechanism, based on the discussion in this paper. 

We consider the expectation value of a local gauge invariant operator, 
\begin{align}
\bbra{\Psi} \lgop\bket{\Psi} 
=\frac {\int{\cal D}\vec\bbeta \int{\cal D}\vec{\tilde\bbeta}
  \psi(\vec\bbeta) \psi(\vec{\tilde\bbeta}) 
  \bra{\vec\bbeta}\psiHlocalbra \lgop \psiHlocaltilde \ket{\vec{\tilde\bbeta}}}
  {\int{\cal D}\vec\bbeta \int{\cal D}\vec{\tilde\bbeta}\psi(\vec\bbeta) \psi(\vec{\tilde\bbeta}) \bra{\vec\bbeta} \vec{\tilde\bbeta}\Bigr\rangle}\,.
\label{eq:regularity_first}
\end{align}
As shown in Eq.~\eqref{eq:homoWF} and Eq.~\eqref{Eq:derivation_locality2}, both $\psi(\vec\bbeta) \psi(\vec{\tilde\bbeta})$ and $\bra{\vec\bbeta}\psiHlocalbra \lgop \psiHlocaltilde \ket{\vec{\tilde\bbeta}}$ are independent of $\bar\zeta^c$. Both the numerator and the denominator on the right-hand side of Eq.~\eqref{eq:regularity_first} contain infrared divergences originating from the integrals over $\bar\zeta^c$. However, since the integrands do not depend on $\bar\zeta^c$, these divergent integrals are canceled between the numerator and the denominator. 

In Refs.~\cite{Urakawa:2010it, Urakawa:2010kr, Tanaka:2012wi, Tanaka:2014ina}, we showed that a non-trivial condition needs to be imposed on the initial quantum state for the IR regularity. If we choose the initial vacuum state inappropriately, the locality condition will not be satisfied, and then IR divergences arise. Here, we show that the condition for the IR regularity discussed in our previous papers is nothing but the locality condition, making our claim in Ref.~\cite{Tanaka:2017nff} more concrete. In the rest of this section, for simplicity, we only consider single-field models of inflation. 

Using Eq.~(\ref{Eq:locality2}), which can be derived from the locality condition (\ref{eq:locality}), we have shown the generalized soft theorem in Sec.~\ref{Sec:ST} and also the cancellation of the IR divergence in this section. In the operator formulation, Eq.~\eqref{Eq:locality2} corresponds to the requirement that ${}^{\raisebox{0.4ex}{{\scriptsize$g$}}}\!{\hat\zeta}^{\rm H}$, which is the dilatation-invariant counterpart of hard modes $\hat\zeta^{\rm H}$, be independent of $\hat\zeta^{\rs} \approx \hat\zeta^{\rm{S(1)}}$, when we perturbatively expand it in the interaction picture. 
Here and hereafter, we count the order of perturbation, putting a superscript $(m)$ with $m=1,\, 2,\, 3,\, \cdots$. 

In what follows, we show that we can choose a solution of $\hat\zeta^{\rm H}$ that satisfies 
\begin{align}
\left[i\hat Q,\, {}^{\raisebox{0.4ex}{{\scriptsize$g$}}}\!\hat\zeta^{\rm H}\right] 
=\frac{\partial \,{}^{\raisebox{0.4ex}{{\scriptsize$g$}}}\!\hat\zeta^{\rm H}}{\partial \hat\zeta^{\rm{S(1)}}(x)}
=0\,. 
\label{eq:CIteration}
\end{align}
For simplicity, we omit the subscript $\alpha$ and ignore irrelevant terms for our discussion, which do not contain $\hat\zeta_{\alpha}^{\rm{S(1)}}$. The Heisenberg equation of motion up to the second order in perturbation is given by 
\begin{align}
  {\cal L}_p \hat\zeta^{\rh (2)}_{\sbp} =\Gamma^{(2)}_{\sbp}(\hat\zeta^{(1)}),
\end{align}
where ${\cal L}_p$ is the linear differential operator that eliminates $\hat\zeta^{(1)}_{\sbp}$, {\it i.e.}, ${\cal L}_p \hat\zeta^{(1)}_{\sbp} = 0$~\footnote{For instance, for a canonical scalar field ${\cal L}_p$ is given by ${\cal L}_p = \partial^2_t + (3 + \varepsilon_2)\dot{\rho} \partial_t + e^{-2 \rho} p^2$ with $\varepsilon_2$ being the second slow-roll parameter.} and $\Gamma^{(2)}_{\sbp}(\zeta)$ represents the quadratic source terms. 
Since $\hat\zeta^{\rs (1)}$ is dominated by the time-independent part, {\it i.e.}, Weinberg's adiabatic mode, in the limit $k\to 0$, the source term in quadratic order can be simply given by $\hat\zeta^{{\rm S(1)}} \Gamma'_{\sbp}\hat\zeta_{\sbp}^{H(1)}$, where $\Gamma'_{\sbp}$ is a differential operator. In other words, we are interested in the part of $\hat\zeta^{{\rm S(1)}}$ that is affected by its constant shift.

Since the equation of motion is also diffeomorphism invariant, it contains $\zeta$ without differentiations only in the combination $e^{-\zeta}\bp$ with the momentum $\bp$. Namely, at the quadratic order, it only appears from the spatial gradient in the d'Alembert operator (a concrete computation can be found, {\it e.g.}, in Ref.~\cite{Tanaka:2012wi}). Therefore, we find 
\begin{align}
 \Gamma'_{\sbp}=\bp\cdot \frac{\partial{\cal L}_p}{\partial \bp}\,. 
\end{align}
Hence, we obtain a formal solution, 
\begin{align}
    \hat\zeta^{\rh (2)}_{\sbp}\approx {\cal L}_p^{-1} \hat\zeta^{\rs (1)}
  \bp\cdot \frac{\partial{\cal L}_p}{\partial \bp}\hat\zeta^{\rh (1)}_{\sbp} +\hat\xi\,,
  \label{eq:zeta2p}
\end{align} 
where $\hat\xi$ represents a homogeneous solution which is eliminated by operating ${\cal L}_p$, {\it i.e.}, ${\cal L}_p \hat\xi=0$. From Eq.~\eqref{eq:zeta2p}, one can obtain $\,{}^{\raisebox{0.4ex}{{\scriptsize$g$}}}\!\hat\zeta^{\rm{H (2)}}$ as 
\begin{align}
    \displaystyle \,{}^{\raisebox{0.4ex}{{\scriptsize$g$}}}\!\hat\zeta^{\rm{H (2)}}=\hat\zeta^{\rm{H (2)}}+\hat\zeta^{\rs (1)} \frac{\partial}{\partial \bp}\cdot \bp \, \hat\zeta^{\rh (1)}_{\sbp}\,. 
\end{align}
Using this solution, the condition~\eqref{eq:CIteration} can be explicitly written down as 
\begin{align}
\frac{\partial\hat\xi}{\partial \hat\zeta^{\rs (1)}}
 =-\left({\cal L}_p^{-1} \bp\cdot \frac{\partial{\cal L}_p}{\partial \bp}
   +\frac{\partial}{\partial \bp}\cdot \bp\right)\hat\zeta^{\rh (1)}_{\sbp}\,,
\end{align}
imposing a non-trivial condition on the homogeneous solution. Since the right-hand is annihilated by acting with ${\cal L}_p$, this equation can be easily integrated to give 
\begin{align}
   \hat\xi = - \hat\zeta^{\rs (1)} \left({\cal L}_p^{-1} \bp\cdot \frac{\partial{\cal L}_p}{\partial \bp}
   +\frac{\partial}{\partial \bp}\cdot \bp\right)\hat\zeta^{\rh (1)}_{\sbp}\,.
\end{align}
One can confirm that this $\hat\xi$ indeed satisfies \begin{align}
    {\cal L}_p \hat\xi =0\,,
\end{align} by using ${\cal L}_{\sbp}\hat\zeta^{\rh (1)}_{\sbp}=0$ and $\partial_t \hat\zeta^{\rs (1)} \simeq 0$. In this way, for an arbitrary choice of $\hat\zeta^{\rh (1)}_{\sbp}$, one can verify Eq.~(\ref{eq:CIteration}) by adjusting the homogeneous solution $\hat\xi$. In Refs.~\cite{Urakawa:2010it, Urakawa:2010kr, Tanaka:2011aj}, we claimed that the adiabatic vacuum satisfies this condition, but we did not show that an arbitrary choice of the initial spectrum is possible. In the special case of the adiabatic vacuum, any additional contribution $\hat\xi$ was not necessary to satisfy this condition, if the integration ${\cal L}_p^{-1}$ is taken from the past infinity by modifying the integration path to converge in the complex plane.

\section{Summary}
In this paper, we have shown that the locality condition~\eqref{eq:locality}, imposed on the quantum state of hard modes, provides a unified criterion for the effective description of inflationary soft modes, generalized soft theorems, the suppression of hard-mode loop corrections, and the infrared regularity of observable correlators.  In particular, the usual consistency relation~\cite{Maldacena:2002vr,Creminelli:2004yq} is recovered when the soft-mode wave function shows only negligible correlation between the curvature perturbation and the other degrees of freedom, including the conjugate momentum of the curvature perturbation itself. 

In Ref.~\cite{Tanaka:2017nff}, it was shown that the locality condition in the form of Eq.~\eqref{eq:localityOld} ensures the soft theorem, the existence of a time-independent solution corresponding to Weinberg's adiabatic mode in the effective dynamics, and the cancellation of infrared divergences.  That analysis was formulated mainly for the case in which the background trajectory is described by a single-field slow-roll model.  However, since Weinberg's adiabatic mode is a solution corresponding to an additive shift of the soft curvature perturbation, the argument can be extended in a rather parallel manner even in the presence of other light fields or a deviation from the slow-roll approximation.  In the present paper, we reformulated the discussion by expanding the quantum state in terms of coherent states, so that the locality condition can manifestly accommodate more general situations, including multiple light fields and non-slow-roll evolution.  Another non-trivial extension is that Weinberg's adiabatic mode is shown to exist as an operation on an arbitrary solution, rather than only as a perturbative solution of the effective dynamics described by the effective action (see Sec.~\ref{SSec:WAM}).

In the last several years, there has been an active debate about the possible enhancement of loop corrections to the curvature perturbation in the superhorizon limit.  Our discussion based on the locality condition~\eqref{eq:locality} provides a concrete criterion to exclude the possibility of such an enhancement.  In particular, as long as the hard-mode state satisfies the locality condition, loop corrections from hard modes are suppressed in the soft limit, and a large contribution that may upset the ordering of the gradient expansion cannot arise from a calculation that respects dilatation symmetry, unless large correlations exist between the soft mode of the curvature perturbation and the other soft modes or the perturbative expansion is invalidated at a small scale.

In this paper, for notational simplicity, we have focused on models with scalar fields and on the adiabatic curvature perturbation.  However, as discussed in Ref.~\cite{Tanaka:2017nff}, the argument can also be generalized to fields with non-zero integer spins.  This generalization slightly modifies the corresponding soft theorem.  Another issue that we have not explicitly discussed is the infrared property of gravitational waves.  As also discussed in Ref.~\cite{Tanaka:2017nff}, the infrared properties of $\zeta$ and gravitational waves share various common aspects.  By considering another large gauge transformation, namely a shear transformation
$x^i \to [e^{S/2}]{}^i_{~j} x^j$
where $S_{ij}$ is a constant traceless tensor, one can develop an argument parallel to the one presented here for $\zeta$ under dilatations.

Finally, let us briefly mention a close relationship between the infrared physics of inflation and that of asymptotically flat spacetime~\cite{Bloch:1937pw, Weinberg:1965nx, Kinoshita:1962ur, Lee:1964is}.  Infrared divergences also appear in gauge theories in asymptotically flat spacetime, such as QED, QCD, and perturbative gravity.  In QED, for example, such infrared divergences are known to be canceled after appropriately dressing unobservable low-energy photons.  Originally, this cancellation was discussed at the level of inclusive cross sections.  Subsequently, Faddeev and Kulish constructed infrared-regular asymptotic states by dressing charged particles with an infinite number of soft photons~\cite{Chung:1965zza, Faddeev:1970te}.  More recently, the relation between soft theorems, asymptotic symmetries, and memory effects has been clarified in the context of large gauge transformations and BMS symmetries
~\cite{Strominger:2013jfa,He:2014cra,Strominger:2017zoo, Kapec:2017tkm}. In our accompanying paper~\cite{ToBeSubmitted_Kuramoto}, we will show that the condition satisfied by the Faddeev-Kulish state corresponds to the dilatation-invariant condition on the quantum state in the context of cosmology.

\acknowledgments
T.~T. and Y.~U. would like to thank Jaume Garriga for the warm hospitality at the University of Barcelona, where a part of this project was conducted. 
T.T. is supported by Grant-in-Aid for Scientific Research under Contract No. JP23H00110, and also by SPIRIT2 2026 of Kyoto University.
Y.~U. is supported by Grant-in-Aid for Scientific Research under Contract No.~JP26H00402(26H00402),  Grant-in-Aid for Scientific Research (B) under Contract No.~JP23K25873 (JP23H01177), and JST FOREST Program under Contract No. JPMJFR222Y.

\appendix

\section{Coherent states}

In this appendix, we summarize basic formulae for coherent states, which were implicitly used in the main text. Let $\hat{x}$ and $\hat{p}$ satisfy (with $\hbar$ being set to 1), 
\begin{align}
    [\hat{x},\hat{p}] = i .
\end{align}
Using a positive parameter $\sigma$, we define
\begin{align}
    \hat{a}
    =
    \frac{1}{\sqrt{2}}
    \left(
        \frac{\hat{x}}{\sigma}
        +
        i\sigma \hat{p}
    \right),
    \qquad
    \hat{a}^\dagger
    =
    \frac{1}{\sqrt{2}}
    \left(
        \frac{\hat{x}}{\sigma}
        -
        i\sigma \hat{p}
    \right),
\end{align}
which satisfy the standard commutation relation for the creation and annihilation operators as 
\begin{align}
    [\hat{a},\hat{a}^\dagger]=1 .
\end{align}
A coherent state $|\beta\rangle$ is defined as an eigenstate of $\hat{a}$ as 
\begin{align}
    \hat{a}|\beta\rangle=\beta|\beta\rangle ,
\end{align}
where $\beta$ is, in general, a complex number.

In the position representation, the coherent-state wave function,
\begin{align}
    \psi_\beta(x)\equiv \langle x|\beta\rangle ,
\end{align}
satisfies
\begin{align}
  \langle x | \hat{a} | \beta \rangle =   \frac{1}{\sqrt{2}}
    \left(\frac{x}{\sigma}+\sigma\frac{\partial}{\partial x} \right)
    \psi_\beta(x) = \beta \psi_\beta(x),
\end{align}
where the conjugate momentum is replaced as $ \hat{p}=-i \partial /\partial x $. Solving this equation, we obtain
\begin{align}
    \psi_\beta(x) = N_\beta  \exp\left[ -\left( \frac{x}{\sqrt{2}\sigma} - \beta  \right)^2 \right].
\end{align}
We determine the normalization constant $N_\beta$ by requiring
\begin{align}
    \int_{-\infty}^{\infty} dx\,|\psi_\beta(x)|^2 = 1,
\end{align}
as 
\begin{align}
    |N_\beta|^2 =  \frac{1}{\sqrt{\pi} \sigma} \exp \left[ \frac{\beta^2}{2} + \frac{\beta^{*\,2}}{2} - |\beta|^2 \right] \,. 
\end{align}
With a phase convention chosen below, the normalized wave function is given by
\begin{align}
    \langle x|\beta\rangle
    = \frac{1}{\pi^{1/4}\sqrt{\sigma}}
    \exp\left[ -\left(\frac{x}{\sqrt{2}\sigma} -
            \beta \right)^2 + \frac{\beta^2}{2} - \frac{|\beta|^2}{2} \right].
    \label{App:coherent_wave function}
\end{align}

The overlap of two coherent states can be evaluated by inserting the resolution of the identity in the position basis,
\begin{align}
    \langle \beta'|\beta\rangle = \int_{-\infty}^{\infty} dx\, \psi^*_{\beta'}(x)\,\psi_\beta(x)=
    \exp\left[-\frac{|\beta|^2}{2} - \frac{|\beta'|^2}{2} + {\beta'}^* \beta \right]
    \label{App:coherent_overlap},
\end{align}
where the Gaussian integral over $x$ is performed. In order to consistently satisfy 
\begin{align}
  \delta(x-x')  = \langle x| x' \rangle =  \frac{1}{\pi} \int d^2\beta\, \langle x'|\beta\rangle
    \langle\beta|x\rangle,
\end{align}
the complete set constructed by the coherent state should be given by
\begin{align}
    \frac{1}{\pi} \int d^2\beta\, |\beta\rangle\langle\beta|  = 1 ,
    \qquad  d^2\beta \equiv d {\rm Re}[\beta]\,d {\rm Im}[\beta] \,. 
    \label{App:coherent_identity}
\end{align}

\section{Validity of approximation in \texorpdfstring{\eqref{eq:KeyApprox}}{Eq.~(X)}}
\label{AppB}
Here, we show that  
\begin{align}
\bra{\Psi} f\!(\hat{\vec\zeta})\,{}^g\!\hat{\cal O}_\alpha\ket{\Psi}
=\int{\cal D}\vec\bbeta \int{\cal D}\vec{\tilde\bbeta}\,
  \psi(\vec\bbeta) \psi(\vec{\tilde\bbeta}) 
  \bra{\vec\bbeta}
  f\!(\hat{\vec\zeta})\,\psiHlocalbra \lgop \psiHlocaltilde 
  \ket{\vec{\tilde\bbeta}}, 
\end{align}
can be approximated as \eqref{eq:KeyApprox}. First, we should notice that 
\begin{align}
 \psi(\vec\bbeta) \psi(\vec{\tilde\bbeta}) 
  \,{}^g\!\hat{\cal O}_\alpha(\bbeta_\alpha,\tilde\bbeta_\alpha)
  =\bra{\Psi}\vec\bbeta\Bigr\rangle 
  {}^g\!\hat{\cal O}_\alpha\Bigl\langle\vec{\tilde\bbeta}\ket{\Psi}
  \equiv \exp\left(-\frac{|\vec\bbeta|^2}{2}-\frac{|\vec{\tilde\bbeta}|^2}{2}\right)\times \hat A\left(\bbeta_\alpha,\tilde\bbeta{}^*_\alpha\right), 
\end{align}
where the last equality defines the operator $\hat A\left(\bbeta_\alpha,\tilde\bbeta{}^*_\alpha\right)$, which includes soft mode operators inherited from $^g\hat{\cal O}_\alpha$. 
The factor $\displaystyle \exp\left(-\frac{|\vec\bbeta|^2}{2}-\frac{|\vec{\tilde\bbeta}|^2}{2}\right)$ comes from the normalization of coherent states $\bra{\vec\bbeta}$ and $\ket{\vec\bbeta}$. 
The function $\hat A\left(\bbeta_\alpha,\tilde\bbeta{}^*_\alpha\right)$, which  is defined excluding this factor, depends only on $\vec\bbeta$ and $\vec{\tilde\bbeta}{}^*$, because the annihilation (creation) operators of soft modes contained in $\bra{\Psi}\vec\bbeta\Bigr\rangle$ $\left(\Bigl\langle\vec{\tilde\bbeta}\ket{\Psi}\right)$ are replaced with $\vec\bbeta$ ($\vec{\tilde\bbeta}$). 

Hence, 
\begin{align}
\int{\cal D}\vec\bbeta \int{\cal D}\vec{\tilde\bbeta}\,
  \psi(\vec\bbeta) \psi(\vec{\tilde\bbeta}) 
  \bra{\vec\bbeta}
  f\!(\hat{\vec\zeta})\,{}^g\!\hat{\cal O}_\alpha(\bbeta_\alpha,\tilde\bbeta_\alpha)\ket{\vec{\tilde\bbeta}}
=\int{\cal D}\vec\bbeta \int{\cal D}\vec{\tilde\bbeta}\,
    \exp\left(-\frac{|\vec\bbeta|^2}{2}-\frac{|\vec{\tilde\bbeta}|^2}{2}\right)
    \bra{\vec\bbeta}f\!(\hat{\vec\zeta}) \hat A\left(\bbeta_\alpha,\tilde\bbeta{}^*_\alpha\right)
    \ket{\vec{\tilde\bbeta}}. 
\end{align}
The last factor can be rewritten as 
\begin{align}
\bra{\vec\bbeta}f\!(\hat{\vec\zeta}) \hat A\left(\bbeta_\alpha,\tilde\bbeta{}^*_\alpha\right)
    \ket{\vec{\tilde\bbeta}}
    \simeq
    f\!({\vec\zeta}) A\left(\bbeta_\alpha,\tilde\bbeta{}^*_\alpha;\btheta{}_{,\alpha}\right)
    \Bigl\langle \vec\bbeta\ket{\vec{\tilde\bbeta}}
=f\!({\vec\zeta}) A\left(\bbeta_\alpha,\tilde\bbeta{}^*_\alpha;\btheta{}_{,\alpha}\right) 
\exp\left(-\frac{|\vec\bbeta|^2}2-\frac{|\vec{\tilde\bbeta}|^2}2+\vec\bbeta{}^*\cdot \vec{\tilde\bbeta} \right)\,,   
\end{align}
where $\beta{}^{a}_{{\rm nad},\alpha}$ are defined as a collection of  
\begin{align}
& \phi^I_\alpha=\frac{\sigma_I}{\sqrt{2}}\left(\beta^{I*}_\alpha +\tilde\beta^{I}_\alpha\right) \,,\cr
& \pi_{I\alpha}=\frac{1}{\sqrt{2}\sigma_I}\left(\beta^{I*}_\alpha -\tilde\beta^{I}_\alpha\right) \,,\cr
\end{align}
excluding $\zeta_\alpha\equiv \phi^0_\alpha$. 
Here, $\simeq$ stands for an approximate equality which holds when we neglect the error from the operator ordering. 
Then, we can rewrite $\tilde\beta^I_\alpha$ included in $\btheta{}_{,\alpha}$ by $-\partial/\partial\tilde\bbeta{}^*_\alpha$ acting on the factor $\displaystyle\exp\left(-\frac{|\vec\bbeta|^2}{2}-\frac{|\vec{\tilde\bbeta}|^2}{2}+\vec\bbeta{}^*\cdot \vec{\tilde\bbeta} \right)$. We perform the integral 
of this factor $\int{\cal D}\vec{\bbeta}\int{\cal D}\vec{\tilde \bbeta}$, taking into account ${\cal D}\tilde \bbeta=\prod_{I,\alpha} (d\tilde\beta^I_\alpha d\tilde\beta^{I*}_\alpha/2\pi i)$. 
Using the integration by parts, these derivative operators can be moved to the other side, and then we can perform the integral $\int d\vec{\tilde\bbeta}$ to obtain 
\begin{align}
   \left[\prod_{I, \alpha} \int \frac{d\tilde\beta^{I}_\alpha}{2\pi i}\right]
   \exp\left(-|\vec{\tilde\bbeta}|^2+\vec\bbeta{}^*\cdot \vec{\tilde\bbeta} \right)
   =\delta(\vec{\tilde\bbeta}{}^*-\vec\bbeta{}^*),   
\end{align} 
and hence, for an arbitrary function $F(\vec\bbeta,\vec{\tilde\bbeta}{}^*;\vec{\bbeta}{}^*)$, we have
\begin{align}
\int{\cal D}\vec{\tilde\bbeta}\, F\!\left(\vec\bbeta,\vec{\tilde\bbeta}{}^*;\vec{\bbeta}{}^*\right)
\exp\left(-|\vec{\tilde\bbeta}|^2+\vec\bbeta{}^*\cdot \vec{\tilde\bbeta} \right) =
F(\vec\bbeta,\vec\bbeta{}^*;\vec\bbeta{}^*)\,. 
\label{eq:identity}
\end{align}
Using this identity, we find 
\begin{align}
\int{\cal D}\vec{\bbeta} \int{\cal D}\vec{\tilde\bbeta}\,\psi(\vec{\bbeta})\psi(\vec{\tilde\bbeta})
& \bra{\vec\bbeta}f\!\left(\hat{\vec\zeta}\right) \hat A\left(\bbeta_\alpha,\tilde\bbeta{}^*_\alpha\right)
    \ket{\vec{\tilde\bbeta}}\cr
 &   \simeq
    \int{\cal D}\vec{\bbeta}\int{\cal D}\vec{\tilde\bbeta}\,
    f\!\left({\vec\zeta}\right) \exp\left(-{|\vec\bbeta|^2}\right)
   \left.A\left(\bbeta_\alpha,\tilde\bbeta{}^*_\alpha;\btheta{}_{,\alpha}\right)
    \right\vert_{\tilde\beta^{I}_\alpha\to -\frac{\overleftarrow\partial}{\partial\beta^{I*}_\alpha}}
   \exp\left(-{|\vec{\tilde\bbeta}|^2}+\vec\bbeta{}^*\cdot \vec{\tilde\bbeta} \right)
    \cr
&  \simeq
    \int{\cal D}\vec{\bbeta}\,
    f\!\left({\vec\zeta}\right) \left.A\left(\bbeta_\alpha,\bbeta{}^*_\alpha;\btheta{}_{,\alpha}\right)
    \right\vert_{\tilde\beta^{I}_\alpha\to\beta^{I}_\alpha} 
\exp\left(-{|\vec\bbeta|^2}\right)\,, 
\end{align}
where we used Eq.~\eqref{eq:identity}, neglecting the terms in which $-\frac{\partial}{\partial\beta^{I*}_\alpha}$ acts on $\beta^{I*}_\alpha$ contained in $A$ in the last equality, which causes only the errors of the same order as the operator-ordering ambiguity. In this way, ignoring this ambiguity, one can show the approximate equality \eqref{eq:KeyApprox}. 

\bibliographystyle{unsrturl}
\bibliography{refst.bib}

\end{document}